\documentclass[preprint,aps,eqsecnum,showpacs]{revtex4}
\usepackage{epsf,graphics,graphicx}
\usepackage{amsmath}
\usepackage{amssymb,latexsym,mathrsfs}

\def\bea{\begin{eqnarray}}
\def\eea{\end{eqnarray}}
\def\ba{\begin{array}}
\def\ea{\end{array}}

\def\beq{\begin{equation}}
\def\eeq{\end{equation}}

\begin{document}

\title{Mutated Hilltop Inflation : A Natural Choice for Early Universe}

\author{Barun Kumar Pal\footnote{Electronic address: {barunp1985@rediffmail.com}},
${}^{}$ Supratik Pal\footnote{Electronic address:
{supratik@isical.ac.in}} ${}^{}$ and B. Basu\footnote{Electronic
address: {banasri@isical.ac.in}} ${}^{}$} \affiliation{Physics and
Applied Mathematics Unit, Indian Statistical Institute, 203
B.T.Road, Kolkata 700 108, India}

\vspace{1in}

\begin{abstract}
We propose a model of inflation  with a suitable potential for a single scalar field which falls in the wide class of hilltop inflation. We derive the analytical expressions for most of the physical quantities related to inflation and show that all of them represent the true behavior as required from a model of inflation.
We further subject the results to observational verification by formulating the theory of perturbations based on our model followed by an estimation for the values of those observable parameters.  Our model is found to be in excellent agreement with observational data. Thus, the features related to the model leads us to infer that this type of hilltop inflation may be a natural choice for explaining the early universe.
\end{abstract}


\maketitle

\section{Introduction}

The inflationary paradigm  for explaining the early universe is in
vogue for quite a few years now. The original motivation for
invoking the idea of inflation \cite{guth} was to resolve the
problems associated with standard big bang cosmology. Coupled with
that is precise observational data from  cosmic microwave background
(CMB) \cite{cmb} observations and other independent measures \cite{nsobs} which have started coming of late, 
thereby enforcing
the models of inflation to face the challenge of passing through the
crucial tests of observations. In the simplest models, inflation is
driven by {\em inflaton}, a single scalar field, which evolves
slowly along a nearly flat potential which is difficult to achieve
in the context of particle physics motivated models
\cite{lyth,liddle}. Further, the initial vacuum quantum fluctuation of the
inflaton is translated into macroscopic cosmological perturbations in the early universe
which have imprints on the CMB radiation.
Thus, any model of inflation should, in principle, not only give a
super-accelerated phase at early time leading to the correct number
of e-foldings but also produce the correct power spectra the imprints of which are
directly observable from CMB observations. As
more and more precise cosmological data is available it is
fascinating to learn from the observations, some crucial clues about
the fundamental physics of the early universe.
Attempts in this
direction from diverse angles include standard particle physics
motivated models \cite{guth,lyth,liddle} and allied phenomenological
models \cite{2, 3, linde,nilles}, string theory inspired models
\cite{stinfl}, inflation from supersymmetry \cite{randall}, warm
inflation  \cite{warm}, multi-field inflation
\cite{multi} and braneworld models \cite{brinfl, pal} among others.  Although most of those models
of inflation have some positive feature or the other, the nature of
field(s) responsible for inflation is still an open question.

Recently a proposal to satisfy the flatness conditions  by considering the inflation occurring near a local
maximum of the potential came out, which is termed as ``hilltop" inflation
\cite{hilllyth1, hilllyth2}. In this paradigm, the cosmological scale leaves the horizon the time the inflaton is on the top of the hill. Nowadays hilltop inflation has turned out to be a very prospective model to explain early universe phenomena in the sense that in these models a flat potential can easily be converted to one with a maximum with the addition of one or two terms from a power law series. Nevertheless, many models can be converted to hilltop by suitable tuning of parameters.

 In this article, we would like to present a variant of the hilltop inflation, the crucial characteristic feature of which is that here the modification to the flat potential is not mere addition of one or two terms from a power  series, rather a hyperbolic function which contains infinite number of terms in the power series expansion, thereby making the theory a more concrete and accurate at the same time. We further notice that our choice of the potential bears close similarity  with its counterpart in mutated hybrid inflation, so far as its variation with the scalar field is concerned. Keeping this in mind, we name our proposal ``mutated hilltop inflation''.
Nevertheless, throughout the article we  succeed in having analytical expressions for the parameters and observables and confront them with observational data. Our model is found to be in excellent agreement with observational data.
We are thus led to believe that mutated hilltop inflation is more or less a natural choice to explain the early universe as well as observations related to perturbations therefrom.

The plan of the paper is as follows : In Section II we propose a model for mutated hilltop inflation and show that for some valid approximation one can indeed have analytical expressions for  the scalar field, scale factor and number of e-foldings. We then prove the validity of the model both analytically and numerically. To this end, we estimate the observable quantities using the slow roll parameters and fit them with observational data. Section III deals with the analysis of the typical energy scale of inflation from our model and shows that the energy scale is consistent with observational bound. In Section IV we study quantum fluctuations based on our model and derive the analytical expressions for the observable quantities related to perturbations, like the power spectra, spectral index  and its running, and the ratio of the tensor to scalar amplitudes. We further subject the results to observational verification by estimating those quantities for our model, and show that our model is in good agreement with observational data as well. We thus succeed in having both analytical expressions and observational consistency of our model. Finally, we end up with some open issues related to our work.

\section{Modeling mutated hilltop inflation}

The potential we would like to propose has the form
\begin{equation}
V(\phi)=V_0 \left[1- {\rm sech}(\alpha\phi) \right]
\label{pot}
\end{equation}
Here $V_0$ represents the typical energy scale for hilltop inflation and  $\alpha$ is a parameter  which has dimension of inverse Planck mass. We will see later on that, in our model, $V_0$ has typical value of
$V_0^{1/4} \sim 10^{16} {\rm GeV}$, which is the  characteristic feature of models based on supergravity theory, consistent with the observational bound as well.
 
 In choosing the above form of the potential we are somewhat motivated by the models of inflation in the framework of supergravity, either in the context of D-branes \cite{dbrane}
or in the paradigm of tachyonic inflation models  \cite{tach}, or  simply in supergravity-inspired models \cite{sugra1}.
The characteristic feature of the earlier models of hilltop inflation is that the inflation occurs near the  maximum of the potential.  The potential (\ref{pot}) as proposed by us  is a generic one having infinite number of terms in  power series expansion which takes into account the two-term approximation incorporated in many of the models of hilltop inflation, {\em e.g.} \cite{hilllyth1, hilllyth2, dutta, oku}. It is worthwhile to mention here that any  potential in supergravity paradigm should have, in principle, infinite number of terms, which is taken care of in the model proposed by us, thereby making the theory  more physically relevant and accurate at the same time. Moreover, our model satisfies the condition of vanishing of the potential and its slope at its absolute minimum \cite{vega} i.e. $V(\phi_{min})=V^{\prime}(\phi_{min})=0$, which characterizes a  significant difference from the  
 the usual hilltop potential. Several models with flat scalar potential are around,  which are based on SUSY and nonlinearly realized symmetry or shift symmetry. But the SUSY alone cannot naturally provide potentials that are flat enough for inflation, once supergravity  effects are included. The realization of a valid mechanism is  through Pseudo Nambu Goldstone Boson (PNGB) \cite{ran1}.  However, the simplest scenario with a single PNGB does not work unless the symmetry breaking scale  is higher than the Planck scale, which is presumably outside the range of validity of an effective field theoretic description.  Here the possible realization of inflaton scale to be higher than the Planck scale is substantiated via a prescription which considers the inflaton as the extra components of gauge fields propagating in extra dimensions \cite{rand1}. As it will turn out in due course, our model results in an inflaton scale higher than the Planck scale, resonating with the above discussions. One may also argue that the effective inflaton potential  in the Minimal Supersymmetric Standard Model (MSSM) \cite{mazumdar} have some similarity (apart from the so-called A-term that characterizes MSSM inflation) with the power law expansion of the potential proposed here.  
Keeping all these points in our mind, we are motivated to choose a novel potential of the said form which have most of the plus points inbuilt.   
Nevertheless, what will turn out is that the form of the potential does have very significant consequences so far as observational aspects related to perturbations are concerned. We will discuss this issue in due course.
Though we do not show a priori how this potential can be realized from an effective field theory, we presume it may be an outcome of extra components of gauge fields propagating in extra dimensions as proposed in \cite{rand1}. 
However, we are yet to make any strong comment on the origin as such. Any progress in this direction will be reported in future.

\begin{figure}[htb]
\centerline{\includegraphics[width=7cm, height=5cm]{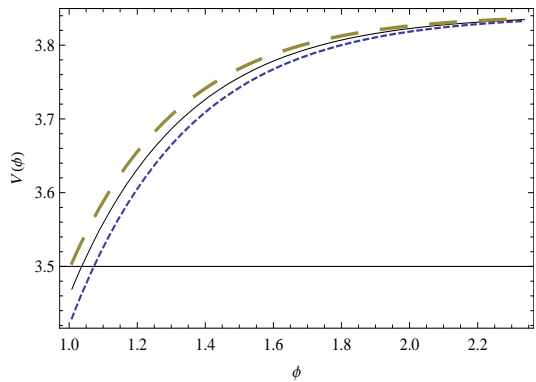}}
  \caption{\label{figVphi} Variation of the potential(~in  units of~$ 10^{-12} M_P^{4}$) with the scalar field. The plots correspond to three sets of values for $\alpha = (2.9, ~3.0, ~3.1) M_P^{-1}$ from  bottom to top.}
\end{figure}
Figure \ref{figVphi} shows the explicit behavior of the potential with the inflaton field for different values of $\alpha$. It will be revealed later on that $\alpha = (2.9 -3.1) M_P^{-1}$ gives the best fit model from observational ground. So, here, and throughout the rest of the paper, we adhere to this range for $\alpha$.
The nature of the potential in Figure \ref{figVphi}
is a characteristic feature of mutated hilltop inflation models, which resemble mutated hybrid inflation models.  
Our model thus falls in the wider class of hilltop inflation, which revisits  the salient features of  inflation from a different perspective.

The initial position of the inflaton field is very important in any model as an initial condition for successful inflation \cite{lin1,lin2,lin3}. The considerations of the naturalness also depend crucially on the underlying assumptions of the model. In the known hilltop potential \cite{hilllyth1}, the inflaton scale is typically 3 orders of magnitude below Planck scale, which  incorporates eternal inflation thereby bypassing  the problem of initial conditions. But, as it appears, this leads to a new  problem related to entropy of the universe. In a bit more details, if the universe is closed, its total entropy must be greater than $10^{9}$ with its total mass at the beginning of the inflation being greater than $10^6 M_P$. At present the entropy of the universe is greater than $10^{87}$. The explanation of this observation requires the assumption that the entropy was extremely large from the very beginning, but that leads to the difficulty of understanding the homogeneity of the large universe. Therefore, though the problem of initial condition can be bypassed, it leads to a new problem which is too important to avoid.
An way of alleviating the initial condition problem for the  low scale inflation is via the consideration of a compact flat or open universe with nontrivial topology \cite{top1,top2,top3,top4,top5} i.e., the compact topologically nontrivial flat or open universes are probable than the standard Friedmann universes for the inflation occurring much below the Planck density. But then one has to explain how it evolved to the Friedmann universe which has to be there after the universe becomes observable. So, if the universe has to boil down to the  observable universe, it is good to rely on Friedmann universe with super-Planckian inflaton scale.   
 
 On the other hand, the models based on chaotic initial conditions, the process of natural inflation occurs at $\phi \geq M_P$ \cite{ran1}. 
From the generic expression for the effective potential 
\begin{equation}
V(\phi)=V_0+\alpha \phi+\frac{m^2}{2}\phi^2+\frac{\beta}{3}\phi^3+\frac{\lambda^4}{4}\phi^4+\sum_n \lambda_n\frac{\phi^{4+n}}{M_P^n} 
 \end{equation}
it may be noted that with the generic assumption $\lambda_n=O(1)$, the effective theory is not under control over the behavior of $V(\phi)$ at $\phi>M_P$  \cite{lyth}. 
 A possible way to overcome the problem of the generic assumption $\lambda_n=O(1)$ is to consider inflation occurring below Planck scale, which creates a new problem, as revealed earlier.
Furthermore, the sub-Planckian inflaton scale also leads to the so called $\eta$-problem in the string  inflation scenarios. In other words, the second slow-roll parameter $|\eta| \equiv M_P^2|\frac{V^{\prime\prime}}{V}|\sim 1$ in string theory while inflation requires $|\eta| << 1$. However,
there are some recent proposals of addressing this problem  \cite{eta1, lyth}. 
It is well-known  that  $\phi>M_P$ attributes to the non-renormalizable quantum gravity with a cut-off at momenta $k \sim M_P$. 
  Thus, the quantum gravity effects are considerable only at super-Planckian inflaton mass scales.
As already discussed in somewhat details, the situation can be dealt with 
\cite{ran1} if $\phi$ is considered as a PNGB  and a scale of spontaneous breaking $f>>M_P$ \cite{ran3,ran2}.
In due course, it will be revealed that our proposal of mutated hilltop inflation deals  with the inflaton scale  $\phi \geq M_P$. This is in accordance with the major conclusion of the above discussions on initial condition leading to super-Planckian inflaton scale. 

The Friedmann equation for the homogeneous and isotropic flat universe, dominated by this type of inflationary potential for the scalar field,
is given by
\begin{equation}\label{fdq}
H^2=\frac{V_0}{3M_P^2}[1-{\rm sech}(\alpha\phi)]
\end{equation}
At this point, one can straightaway use the slow roll parameters to analyze the outcome of the above equation. Instead, we will use the Hubble slow roll parameters which are defined as
\begin{equation}\label{eta}
\epsilon_H=2M_P^2~ \left(\frac{1}{H} \frac{dH}{d\phi}
\right)^2,~~~~~ \eta_H=2M_P^2~ \left(\frac{1}{H} \frac{d^2 H}{d
\phi^2}\right)
\end{equation}
Our basic intention of using Hubble slow roll parameters is not to restrict ourselves in the usual slow roll approximation which are somewhat limited in a generic supergravity theory \cite{sugra1} but to use a more accurate version of the same given by the above Hubble slow roll parameters.
Consequently, with the above Friedmann equation for the typical inflaton potential (\ref{pot})
the  Hubble slow roll parameters take the form
 \begin{eqnarray}
\epsilon_H &=& \frac{M_P^2}{2}~\frac{\alpha^2 {\rm
sech}^2(\alpha\phi)\tanh^2(\alpha\phi)}{[1- {\rm
sech}(\alpha\phi)]^2}
\label{epsh} \\
\eta_H &=& M_P^2 \frac{\alpha^2 {\rm sech}(\alpha\phi)[{\rm
sech}^2(\alpha\phi)-\tanh^2(\alpha\phi)]}{[1-{\rm sech}
(\alpha\phi)]}- \frac{M_P^2}{2}\frac{\alpha^2 {\rm
sech}^2(\alpha\phi)\tanh^2(\alpha\phi)}{[1-{\rm
sech}(\alpha\phi)]^2}
 \label{etah}
\end{eqnarray}

The equation supplementary  to (\ref{fdq}), {\em i.e.}, the Klein-Gordon equation for the homogeneous scalar field in the cosmological background is given by
\begin{equation}\label{kg}
\ddot \phi +3H \dot \phi+\frac{dV}{d\phi}=0
\end{equation}
where an overdot denotes a derivative with respect to time.
Imposing the  slow-roll approximation $|\eta_H|\ll 1$ and
$|\epsilon_H| \ll 1$,
 and using our potential (\ref{pot}) the evolution equation (\ref{kg}) for the scalar field boils down to
 \begin{equation}
 \frac{\sqrt{1- {\rm sech}(\alpha\phi)}}{{\rm sech}(\alpha\phi) \tanh (\alpha\phi)} d\phi+\alpha\sqrt{\frac{V_0}{3}}M_P~dt=0
\label{kgpot}
 \end{equation}

An exact solution for the above equation can indeed be obtained by direct integration. Written explicitly, the solution looks
 \begin{eqnarray}
 \frac{1}{\sqrt{2}\alpha}[-\sqrt{2}\sinh^{-1}(\sqrt{2} \sinh\frac{\alpha\phi}{2})+
 2(\tanh^{-1}[\frac{\sinh(\frac{\alpha\phi}{2})}{\sqrt{\cosh(\alpha\phi)}}] \nonumber \\
 +\sqrt{\cosh(\alpha\phi)} \sinh(\frac{\alpha\phi}{2}))]=-\alpha\sqrt{\frac{V_0}{3}}M_P t+ {\rm constant}
 \end{eqnarray}

However, though exact, this solution is not very useful for
practical purpose because of its complicacy. Precisely, in
order to deal with estimates of observable quantities, one has to
use this expression for further analytical and numerical
calculations, which is not easy to handle. A  route bypassing the
problem is to do numerical estimation for the parameters without
bothering much about the analytical expressions as such. Instead,
one can search for an approximate analytic solution for Eq
(\ref{kgpot}) which will help us derive analytical expressions for
most of the parameters involved with the theory of inflation  and
perturbations therefrom. This will, in turn, help us visualize the
pros and cons of the scenario both analytically and numerically
for quantitative estimation at a later stage. We foresee more
merit in this second route and will follow this subsequently.

In order to obtain the analytical expressions for the parameters we  incorporate the following steps.
 Keeping the terms upto the second power of ${\rm sech}(\alpha\phi)$ in Eq (\ref{kgpot}), the analytic solution for the equation is found to be
 \begin{equation}
 \sinh(\alpha\phi)=-\alpha^2\sqrt{\frac{V_0}{3}}M_P~t+ {\rm constant}
 \end{equation}
where the constant can be found from the condition that
at the end of inflation $t=t_{end}$, the scalar field has the value $\phi=\phi_{end}$. This  readily gives
 \begin{equation}\label{cond}
 \sinh(\alpha\phi)=\alpha^2\sqrt{\frac{V_0}{3}}M_P(d-t)
 \end{equation}
 with
 \begin{equation}
 d=t_{end}+\frac{\sinh(\alpha\phi_{end})}{\alpha^2\sqrt{\frac{V_0}{3}}M_P}
 \end{equation}
which can be estimated once $\phi_{end}$ is known from observational bound for the Hubble slow roll parameters at the end of inflation.

\begin{figure}[htb]
 \centerline{\includegraphics[width=7cm, height=5cm]{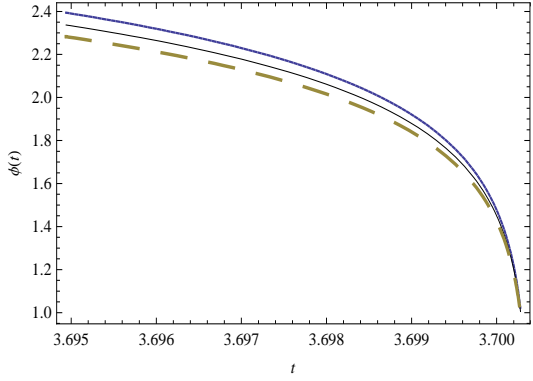}}
  \caption{\label{figphit} Variation of the inflaton field with time~(in units of $10^{10}M_P^{-1}$) for the same set of values for $\alpha$}
\end{figure}

In Figure \ref{figphit} we show the variation of the scalar field with time for the best fit value of $\alpha = (2.9 - 3.1) M_P^{-1}$. The plots clearly show that the scalar field gradually decays as it approaches towards the end of inflation, finally reaching a value $\phi_{end}$ as determined above.

With the expression (\ref{cond}) for the scalar field as a
solution for  the Klein-Gordon  equation in the slow-roll regime,
we arrive at the following equation
\begin{equation}
\frac{da}{a}=\frac{\alpha^2 \frac{V_0}{3}(d-t) }{\sqrt{1+\alpha^4
\frac{V_0}{3}M_P^2(d-t)^2}}dt
\end{equation}
Consequently, the solution for the scale factor turns out to be
\begin{equation}\label{scale}
a(t)=a_1~\exp \left[-(\alpha
M_P)^{-2}\sqrt{1+\alpha^4\frac{V_0}{3}M_P^2(d-t)^2} \right]
\end{equation}
where
\begin{equation}
a_1=a(t_{end})~\exp[(\alpha M_P)^{-2}\cosh(\alpha\phi_{end})]
\end{equation}
is the scale factor at the end of inflation, scaled  by the
exponential term.


\begin{figure}[htb]
 \centerline{\includegraphics[width=7cm, height=5cm]{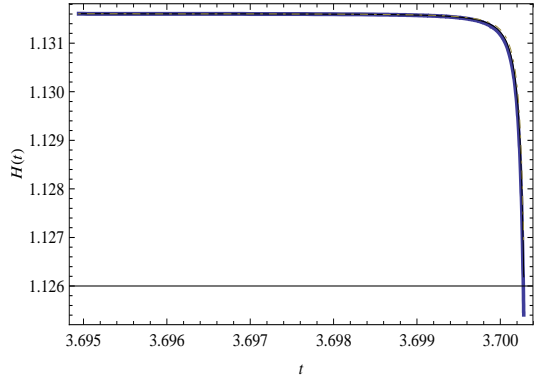}}
  \caption{\label{figHt} Variation of the Hubble Parameter~(in units of ~$10^{-6} M_P$)  with time~(in units of $10^{10} M_P^{-1}$) for the best fit model}
\end{figure}

In Figure \ref{figHt} we show the variation of the Hubble parameter with time as calculated from the above expression for the scale factor. The plot represents the typical characteristic of the Hubble parameter during inflationary phase.

Further, from observational ground it is also interesting to find out the expressions for
the number of e-foldings  which is defined as
\begin{equation}
N=\ln \frac{a(t_{end})}{a(t_{in})}=\int_{t_{in}}^{t_{end}}H~dt
\end{equation}
For our model, the expression for the number of e-foldings turns out to be
 \begin{eqnarray}
 N&\sim & M_p^{-2}\int_{\phi_{in}}^{\phi_{end}}\frac{V}{V^\prime}d\phi\\
       &=& (\alpha^2M_p^2)^{-1}[\cosh(\alpha\phi)-\ln \cosh^2(\frac{\alpha\phi)}{2})]_{\phi_{end}}^{\phi_{in}}
 \end{eqnarray}

\begin{figure}[htb]
 \centerline{\includegraphics[width=7cm, height=5cm]{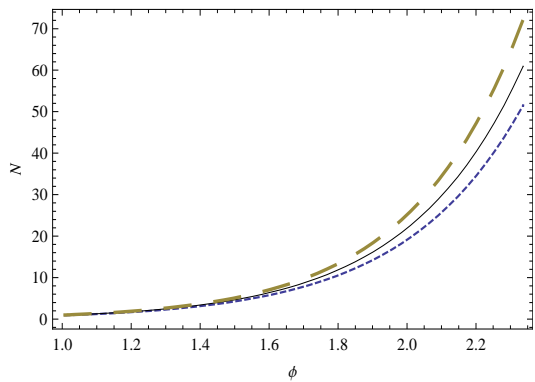}}
  \caption{\label{figNphi}Plot of the number of e-foldings versus scalar field for three sets of values  for $\alpha$}
\end{figure}

Figure \ref{figNphi} shows the plot of e-foldings versus scalar field for the best fit values of $\alpha$. The figure is in agreement with the observational requirement for e-foldings  $56 \leq N \leq 70$.

In Table \ref{tab1} we have estimated the values of different observables form our model using slow roll parameters for three different sets for the values of $\alpha = 2.9, ~ 3, ~3.1$ which are found to give best fit results.

 \begin{table}[htb]
\begin{tabular}{|c|c|c|c|c|c|c|c|c|}
\hline $\alpha$  & $\epsilon_H<1$ &$|\eta_H|<1$ & $\phi_{end}$ &
$\phi_{in}$ & N & $\phi_{hc}$
& $n_s$ & r  \\
$M_P^{-1}$&$\phi\geq M_P$ &$\phi\geq M_P$& $M_P$ & $M_P$& & $M_P$ & &  \\
\hline
&&& & 2.44625 & 70 & 2.3943 &
0.96738 & 2.5$\times 10^{-4}$
\\
2.9 & 0.59886 & 1.02192& 1.02192 & 2.39431 & 60 & 2.3331 & 0.96101 & 3.6 $\times 10^{-4}$   \\
& & & & 2.37111 & 56 & 2.3052 & 0.95771 & 4.2 $\times 10^{-4}$   \\
\hline
&&&& 2.38713 & 70 & 2.33689
& 0.9674 & 2.35 $\times 10^{-4}$  \\
3.0& 0.58759 & 1.00796 & 1.00796  & 2.33689 & 60 & 2.27769 & 0.96099 & 3.36 $\times 10^{-4}$ \\
& & & & 2.31446 & 56& 2.25071 & 0.95769 & 3.9 $\times 10^{-4}$  \\
\hline
&&& & 2.33112 & 70 & 2.28248 & 0.96736
& 2.2 $\times 10^{-4}$
\\
3.1 & 0.57681 & 0.99435 & 0.99435 & 2.28248 & 60 & 2.22516 & 0.96099 & 3.1 $\times 10^{-4}$   \\
& & & & 2.26076 & 56 & 2.19903 & 0.95768 & 3.7 $\times 10^{-4}$  \\
\hline
\end{tabular}
\caption{Table for different parameters related to the present model of inflation as calculated from the slow roll parameters}
\label{tab1}
\end{table}

The salient features of our model for inflation  worth discussing at this point.
\begin{itemize}
 \item The table reveals that the second Hubble show roll parameter $\eta_H$ gives the true bound for scalar field at the end of inflation $\phi_{end}$ which has been used to find out its value $\phi_{in}$ when the inflation begins for three different values of e-foldings  $N$ within the observational bound $56 \leq N \leq 70$.

\item The cosmological scale leaves the horizon during about 10 e-foldings. The values of the scalar field during the horizon crossing $\phi_{hc}$ for three different values of  $\phi_{in}$ has been estimated.

\item We have also calculated the two crucial observable
parameters related to perturbation in this model, namely, the
spectral index $n_s = 1+2 \eta_H-4\epsilon_H$ and the ratio of the
tensor to scalar amplitudes $r < 0.002 (\Delta \phi/M_p)^2
(60/N)$. The observational bound to these parameters are given by
the CMB anisotropy \cite{cmb} and other independent probe
\cite{nsobs} to be $0.948 < n_s <1$ and   $r \leq 0.002$. In Table \ref{tab1} we
show from our model that the values for those observable parameters
are well within the observational bounds. Our model thus fits
well with observations.
\end{itemize}

Thus the mutated hilltop inflation model turns out to be a natural choice for explaining the early universe. We will establish this claim more strongly in subsequent sections.
A crucial point to note here is that the values of the parameters related to perturbation are calculated using slow roll parameters. In Section-IV we will perform a more accurate technique of calculating those parameters by a rigorous development of the theory of perturbation based on this model, which will further prove the credentials of the model.


\section{Analysis of the energy scale}

Let us now analyze the typical energy scale for inflation in this mutated hilltop inflation model and check for its consistency with supergravity framework.
In order to do that let us get back to Eq (\ref{cond}) to have
\begin{equation}
\cosh(\alpha \phi) \approx \sqrt{1+\alpha^4\frac{V_0}{3} M_P^2 (d-t)^2}
 \approx \alpha^2\sqrt{\frac{V_0}{3}}M_P (d-t)
\end{equation}
with the approximation valid for the values of the parameters used in the theory,
resulting in
\begin{equation}
\tanh(\alpha \phi) \approx 1
\end{equation}
Consequently, the second Hubble slow roll parameter given by Eq (\ref{etah}) boils down to
\begin{equation}
\eta_H =-\left( \frac{M_P^2\alpha^2}{2} \right)  \frac{2\alpha^2\sqrt{\frac{V_0}{3}}M_P(d-t)-1}
{\left(\alpha^2 \sqrt{\frac{V_0}{3}}M_P(d-t)-1\right)^2}
\end{equation}
Substituting the variable
\begin{equation}
\alpha^2 \sqrt{\frac{V_0}{3}}M_P(d-t)-1=y
\end{equation}
we get a quadratic equation for $y$
\begin{equation}
2|\eta_H| y^2-2M_P^2\alpha^2 y-M_P^2\alpha^2=0
\end{equation}
which has the physically relevant solution
\begin{equation}
y=\frac{M_P^2\alpha^2+ \sqrt{M_P^4\alpha^4+2M_P^2\alpha^2|\eta_H|}}{2|\eta_H|}
\end{equation}
Thus the expression for the energy scale of inflation turns out to be
\begin{equation}
V_0^{1/4}=3^{1/4}\left[ \frac{2|\eta_H|(\sinh(\alpha\phi_{end})+1)+M_P^2\alpha^2+
\sqrt{M_P^4\alpha^4+2M_P^2\alpha^2|\eta_H|}}{2\alpha^2M_P|\eta_H|(t_{end}-t)}\right]^{1/2}
\label{V0}
\end{equation}

\begin{figure}
 \centerline{\includegraphics[width=7cm, height=5cm]{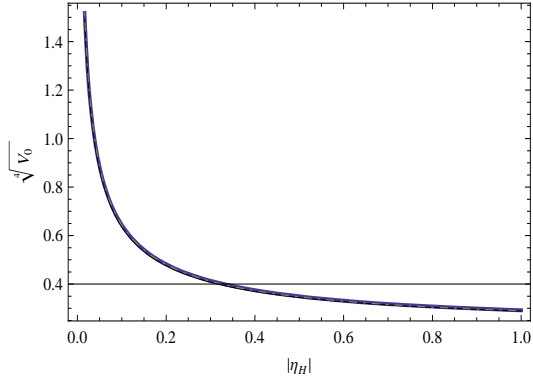}}
  \caption{\label{figVeta} Energy scale for inflation $V_0^{1/4}$ (in units of ~$10^{16}{\rm GeV}$)~versus slow roll parameter $\vert \eta_H \vert$}
\end{figure}
In Figure \ref{figVeta} we have plotted the energy scale for
inflation $V_0^{1/4}$ versus slow roll parameter $\vert \eta_H
\vert$ taking the value of $t$ at the time of horizon crossing.
The plot reveals that in this paradigm the universe had undergone
a slow roll followed by a fast roll towards the end of inflation.
This is a characteristic feature of our model which makes it
distinct from several other models in the same vein.

Further, from the expression (\ref{V0}) it can be verified that the observational bound for $V_0^{1/4} < 2.71 \times 10^{16} {\rm GeV}$ is satisfied in our model for the allowed range of the slow roll parameter so that typical energy scale of inflation, as obtained from our model, is $V_0^{1/4} \sim 10^{16} {\rm GeV}$. Thus, our model shows remarkable consistency with supergravity framework and observational bound as well.


\section{Quantum fluctuation and observable parameters}

We will now concentrate on the theoretical and observational aspects related to perturbations in our model. As proposed in a series of papers \cite{qfluc, pert, langpert} the initial vacuum
quantum fluctuation is transformed by the inflated expansion to macroscopic cosmological perturbations. The latter is responsible for scalar and tensor perturbations directly related to observations. In what follows we will employ the theory of quantum fluctuation derived from our model followed by metric based perturbations to find out the power spectrum -- which will directly relate the quantum fluctuations to observables related to classical perturbations, such as the spectral index and its running, and the ratio of tensor to scalar amplitudes.  The results will then be subject to confrontation with observations, such as the data obtained from CMB \cite{cmb}, in order to show the validity of our model from observational ground.

 Before going into the details of perturbations it  worths noting that in the expression for the scale factor (\ref{scale}), the term $\alpha^4\frac{V_0}{3}M_P^2(d-t)^2$ is much larger than unity during inflation, so we have
\begin{equation}
a(t)\sim a_1~\exp[ M_P^{-1}\sqrt{\frac{V_0}{3}}(t-d)]
\end{equation}
Hence the scale factor behaves pretty close to de Sitter so that we can use near-de Sitter approximation wherever necessary. We shall employ this argument in the following analysis.


\subsection{Curvature perturbation}

Using the above near-de Sitter approximation and defining a variable
$v = a(\eta) \phi$
where $\eta$ is the conformal time (which is negative consistent with the positive scale factor), the new scalar field $v$ can be quantized by expanding it in Fourier space and  the corresponding quantum field is given by
\begin{equation}
 \hat v(\eta, \overrightarrow x) = \frac{1}{(2 \pi)^{3/2}} \int d^3k
\left[\hat a_k v_k(\eta) e^{i \overrightarrow k.\overrightarrow x}
+\hat a_k^{\dagger} v_k^*(\eta) e^{-i \overrightarrow k.\overrightarrow x} \right]
\end{equation}
where the creation and annihilation operators $\hat a_k^{\dagger}$ and $\hat a_k$ satisfy the usual commutation relations.

The equation of motion for the $k$-th Fourier mode of the quantum field $\hat v$ is given by its classical analogue
\begin{equation}\label{veeke}
v_k^{\prime\prime}+\left(k^2-\frac{z^{\prime\prime}}{z}\right)v_k=0
\end{equation}
where the field $v$ is related to the comoving curvature
perturbation $R$ by $v=-zR$ and $z=\frac{a \phi^\prime}{{\cal{H}}}$
and a prime denotes a derivative with respect to conformal
time.
In our model
\begin{equation}
z \approx \alpha^{-1}M_P\sqrt{\frac{3}{V_0}}\vert\eta\vert^{-1}\left[\ln \left(a_1M_P^{-1}\sqrt{\frac{V_0}{3}} \vert\eta\vert \right) \right]^{-1}
\end{equation}
In the slow roll inflation the evolution of $\phi$ and $H$ are much slower than that of the scale factor $a$. So
we get the relation
$$\frac{z^{\prime\prime}}{z} \approx \frac{a^{\prime\prime}}{a}$$
This reduces  Eq (\ref{veeke})
to
\begin{equation}\label{veta}
v_k^{\prime\prime}+\left(k^2-\frac{2}{\eta^2}\right)v_k=0
\end{equation}
The general solution for which is  given by
\begin{equation}\label{vee}
v_k=c_1~\exp(-ik\eta)\left(1-\frac{i}{k\eta}\right)~+~c_2~\exp(ik\eta)\left(1+\frac{i}{k\eta}\right)
\end{equation}
From the normalization condition $\langle v_k^* ,v_k\rangle=1$ and
$\lim_{\eta\rightarrow -\infty}
v_k=\frac{\exp(-ik\eta)}{\sqrt{2k}}$ we get
$c_1=\frac{1}{\sqrt{2k}}$ and $c_2=0$ such that we are finally
left with the following expression for $v_k$
\begin{equation}\label{veek}
v_k=\sqrt{\frac{1}{2k}}\exp(-ik\eta)\left(1-\frac{i}{k\eta}\right)
\end{equation}

Once the above expression has been obtained,
the power spectrum for the comoving curvature perturbation can readily be obtained from the relation
\begin{equation}
P_R(k)=\frac{k^3}{2\pi^2}|R_k|^2=\frac{k^3}{2\pi^2}\frac{|v_k|^2}{z^2}
\end{equation}
Consequently, in our model, the power spectrum turns out to be
\begin{equation}\label{ps}
P_R(k)=\frac{\alpha^2V_0}{12\pi^2 M_P^2}(1+k^2\eta^2)\left[\ln
\left(a_1M_P^{-1}\sqrt{\frac{V_0}{3}}\vert\eta\vert
\right)\right]^{2}
\end{equation}
Now, since  at the time of horizon crossing, $k=aH=-\eta^{-1}$, the above expression reduces to
\begin{equation}\label{pshc}
P_R|_{k=aH}=\frac{\alpha^2V_0}{6\pi^2 M_P^2}\left[\ln
\left(a_1M_P^{-1}\sqrt{\frac{V_0}{3}}\vert\eta\vert
\right)\right]^{2}
\end{equation}
With the above expression for power spectrum, one can  obtain the
expression for the scalar spectral index which is given by
\begin{equation}
n_s=1+ \frac{d \ln P_R(k)}{d \ln k}|_{k=aH}=1-2\left[\ln
\left(a_1M_P^{-1}\sqrt{\frac{V_0}{3}}\vert\eta\vert
\right)\right]^{-1}
\end{equation}
As discussed earlier, the spectral index should satisfy the observational bound $0.948 < n_s < 1$, which was earlier shown to satisfy in our model from slow roll parameters. Secondly, the parameter $P_R^{1/2}$ has the observational bound from CMB fluctuations \cite{cmb} as $P_R^{1/2} \sim 5 \times 10^{-5}$. We will estimate these observable parameters from the above analysis and validate our model from observational ground.

Further, WMAP3 datasets \cite{wmap3} indicate towards the running of
the spectral index so that the spectral index is not strictly scale
invariant, which will, in turn, have imprints on the CMB spectrum.
Thus a nonzero value for this parameter serves as a crucial
observational test for any model of inflation. We have also
succeeded in calculating the running of the spectral index which
turns out in our model to be
\begin{equation}
\displaystyle \frac{dn_s}{d \ln k}{\displaystyle \left|_{k=aH}
\right.}=-2 \left[\ln
\left(a_1M_P^{-1}\sqrt{\frac{V_0}{3}}\vert\eta\vert
\right)\right]^{-2}
\end{equation}
Clearly, the quantity within the parenthesis is nonzero resulting in a non-vanishing value for the running  of the spectral index. 

\begin{figure}[htb]
 \centerline{\includegraphics[width=7cm, height=5cm]{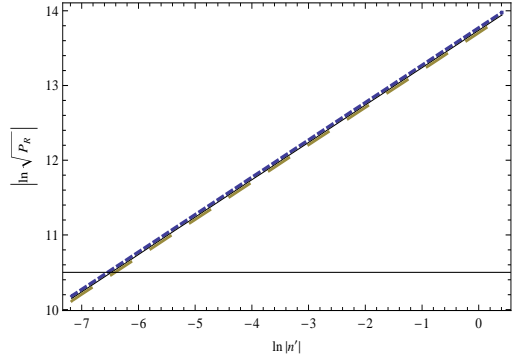}}
  \caption{\label{figPrn} Variation of the square root of power spectrum  versus running in logarithmic scale}
\end{figure}

Figure \ref{figPrn} shows the variation of  the logarithm of $P_R^{1/2}$ with the logarithm of the absolute value for the running of spectral index. The plot shows a straight line which is quite apparent from the analytical expressions.

\begin{figure}[htb]
 \centerline{\includegraphics[width=7cm, height=5cm]{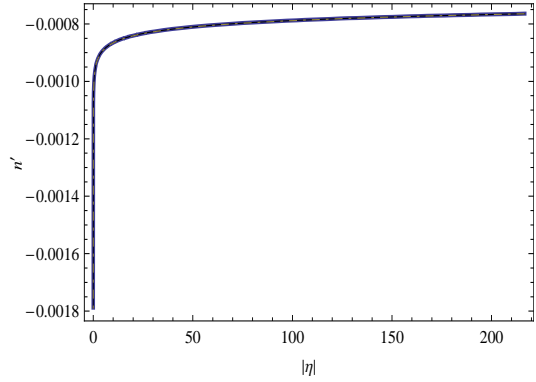}}
  \caption{\label{figrneta} Running of spectral index versus conformal time}
\end{figure}

In Figure \ref{figrneta} we plot the running of spectral index
versus conformal time. The plot shows that the running saturates
to an absolute value $8 \times 10^{-4}$ after a certain value of
the conformal time, which shows a tiny nonzero value for this parameter, thereby validating our model from this significant observational test from WMAP3 \cite{wmap3}.


\subsection{Tensor fluctuation and gravitational waves}

Another interesting observable feature is the possibility of having primordial gravitational waves form tensor fluctuations which also serve as a crucial test for any theory of inflation. As already mentioned, the ratio of tensor to scalar amplitudes should satisfy the observational bound $r \leq 0.002$, which has been shown to satisfy in our model based on slow roll parameters. Here we shall derive it more accurately from the first principle of tensor fluctuation.

The fluctuation equation for the tensor amplitudes is given by
\begin{equation}\label{hk}
h_k^{\prime\prime}+2{\cal{H}}h_k^{\prime}+ k^2 h_k =0
\end{equation}
With the substitution $h_k=\frac{\sqrt{2}}{M_P}\frac{u_k}{a}$ the above equation boils down to the following equation for $u_k$
\begin{equation}\label{uh}
u_k^{\prime\prime}+\left(k^2-\frac{a^{\prime\prime}}{a}\right)u_k=0,
\end{equation}
 which is exactly the same as for $v_k$ in (\ref{veta}), albeit now for the $k$-th Fourier mode for tensor fluctuation, which makes the scenario distinct from Eq (\ref{veta}) from observational ground.
The solution for the above equation can readily be written as
\begin{equation}\label{ukh}
u_k=\sqrt{\frac{1}{2k}}\exp(-ik\eta)\left(1-\frac{i}{k\eta}\right)
\end{equation}
Thus, the power spectrum  $P_{h_k}$ for the $k$-th mode of tensor fluctuation is given by
\begin{equation}\label{phk}
P_{h_k}=\frac{V_0}{6\pi^2M_P^4}(1+k^2\eta^2)
\end{equation}
from where one can have the dimensionless power spectrum of the tensor fluctuation
\begin{equation}\label{pt}
P_{T}=2 P_{h_k}=\frac{V_0}{3\pi^2M_P^4}(1+k^2\eta^2)
\end{equation}
so that we finally arrive at its value during horizon crossing
\begin{equation}\label{ptf}
P_{T}|_{k=aH}=\frac{2V_0}{3\pi^2M_P^4}
\end{equation}
Thus, the expression for the ratio of tensor to scalar amplitudes, given by the ratio of the corresponding power spectra, turns out to be
\begin{equation}
 r = \frac{P_{T}|_{k=aH}}{P_R|_{k=aH}} = \frac{4}{\alpha^2 M_P^2} \left[\ln \left(a_1M_P^{-1}\sqrt{\frac{V_0}{3}}\vert\eta\vert \right)\right]^{-2}
\end{equation}
which we will use direct in calculating the above observable quantity for our model and subject it to  observational verification.

In Table \ref{tab2} we estimate the observable parameters from the first principle of the theory of fluctuation as derives in this section for three sets of values of $\alpha$. For estimation, we take the following representative values for the quantities involved (for $N=60$):
$V_0^{1/4}=1.4\times10^{-3}M_P$,
$ t_{end}\simeq3.70028\times10^{10}M_P^{-1}$,
$t_{in}\simeq3.69493\times10^{10}M_P^{-1}$ and
$a_{in}\simeq1.85184\times10^{-1}M_P^{-1}$.
Also, we have assumed that the cosmological scale leaves the horizon
during first 10 e-foldings to get:
$ d\simeq3.70038\times10^{10}M_P^{-1}$,
$ \eta_{in}\simeq-4.77201\times10^6$ and
$\eta_{end}\simeq-4.17816\times 10^{-20}$.
Here ``hc" represents the value during horizon crossing.

\begin{table}
\begin{tabular}{|c|c|c|c|c|c|c|}
\hline $\alpha $ & $a_1$ & $t_{hc}$ &  $|\eta_{hc}|$ & $P_R^{1/2}$&$n_{s}$& $r$ \\
$M_{P}^{-1}$&$M_{P}^{-1}$&$M_{P}^{-1}$ & &  & &  \\
 \hline $2.9$& $6.7091\times10^{25} $
 &$3.69582\times10^{10}$&$216.649$&$3.7784\times10^{-5}$&0.9609 &$ 1.8176\times10^{-4}$\\
\hline $3.0$& $6.6492\times10^{25} $
 &$3.69582\times10^{10}$&$216.649$&$3.9080\times10^{-5}$&0.9609 &$ 1.6991\times10^{-4}$\\
\hline $3.1$& $6.5945\times10^{25} $
 &$3.69582\times10^{10}$&$216.649$&$4.0377\times10^{-5}$&0.9609 &$ 1.5917\times10^{-4}$\\
 \hline
\end{tabular}
\caption{Table for the observable quantities as obtained from the theory of fluctuations}
\label{tab2}
\end{table}

The table shows remarkable coincidence with the results obtained in Table 1 using slow roll parameters. At the same time, it gives a more accurate result and succeeds to  a great extent so far as observational features of the model is concerned. From the table it is quite clear that the observable parameters related to perturbations, {\em viz}, $P_R^{1/2}$, $n_{s}$ and $r$, as calculated from our model, are in excellent agreement with observational bound.

The spectrum $P_{T}|_{k=aH}$ of tensor  perturbation  conveniently specified by the
tensor fraction $r= \frac{P_{T}|_{k=aH}}{P_R|_{k=aH}}$ yields the relation $r=-8n_T$ in the slow-roll approximation \cite{andrew,david1,david2,lin4}. However, one should  note that the usual relation 
may not strictly hold  for our model. The interpretation for this conclusion is as follows. The usual practice in calculating the relation  between the observable quantities $r$ and  $n_T$ is to use the de Sitter result $aH = - \eta^{-1}$.  However, this is  strictly valid for a de Sitter universe only. 
As a matter of fact, most of the models fail to obtain analytical expressions, and, hence, subject the result $r=-8 n_T$ directly to observational verification. In this article we followed the same prescription in calculating the quantities and confronting them with observations. 
Of course, observationally one can not go too far from de Sitter, but the analytical results,  from a model other than de Sitter, may lead to a relation not exactly identical to, but pretty close to $r=-8 n_T$, at least observationally.
 Strictly speaking, in order to get a relation between these two observable quantities one has to reformulate the perturbation theory from exact expression of scale factor as obtained from the  model, without using the  de Sitter result $aH = - \eta^{-1}$ a priori. Thus, a rigorous  calculation alleviating the de Sitter relation may lead to a modified version of the relation $r=-8 n_T$. Fortunately, we do have analytical expressions in our model. We have now engaged ourselves in obtaining this modified relation analytically.   
Some work in this direction is in progress \cite{ourpert} and we have found some interesting results. We hope to report this in near future.

\section{Summary and outlook}

In this article we have proposed a variant of hilltop inflation models, called mutated hilltop inflation, driven by a hyperbolic potential for the scalar field that has intriguing feature of producing analytical expressions for most of the quantities. To this end, we have derived the expressions for the scalar field, the scale factor, the number of e-foldings and the typical energy scale for inflation from our model. The results have then be subject to observational tests by finding out the values of observable quantities from slow roll parameters. Next, we have  engaged ourselves in formulating the theory of quantum fluctuation in our specific model and have used the results to find out the power spectra for both scalar and tensor fluctuations. These expressions for power spectra have then been used to find out the most crucial observable parameters {\em i.e.}, spectral index, the ratio of tensor to scalar amplitudes and the running of spectral index. All these parameters have  been evaluated during horizon crossing. The results match with their counterparts as calculated from slow roll parameters very well and show excellent agreement with CMB data and other independent observations. We thus infer that mutated hilltop inflation is more or less a natural choice for the explanation of early universe phenomena.

Certain features still remain as open issues in this model the most crucial of them being a more rigorous development of the perturbation theory in this framework. The quantum fluctuations have been studied using near-de Sitter approximation. Alleviating this approximation would definitely alter the analytical expressions, though, predictably, the numerical results will not change significantly. However, a reformulation of the theory of fluctuations with the exact expression for scale factor is useful to have important physical insight and check with more and more accurate data available.  This is  a rather formidable task because of the complications arising in having analytical expressions. The work in this direction will be reported shortly \cite{ourpert}.


\section*{Acknowledgments}

We thank S. Das and S. Ghosh for useful discussions.
BKP thanks Council of Scientific and Industrial Research, Govt. of
India for financial support through Junior Research Fellowship
(Grant No. 09/093 (0119)/2009).


\end{document}